\documentclass[a4paper,fleqn,usenatbib]{mnras}
\usepackage{threeparttable}
\makeatletter
\newlength{\abovecaptionskip}%
\makeatother
\usepackage{threeparttable}%

\usepackage[T1]{fontenc}
\usepackage{ae,aecompl}
\usepackage[normalem]{ulem}

\usepackage{graphicx,soul}	
\usepackage{amsmath}	
\usepackage{amssymb}	

\title[Outer disk asymmetrical motions]
{3D Asymmetrical motions of the Galactic outer disk with LAMOST K giant stars}

\author[H. F. Wang et al.]
{Haifeng Wang,$^{1,2}$\thanks{E-mail:
hfwang@bao.ac.cn} 
Mart\'\i n L\'opez-Corredoira,$^{3,4}$
Jeffrey L. Carlin,$^{5}$ 
Licai Deng$^{1,2}$
\newauthor 
\\
$^{1}$Key Lab of Optical Astronomy, National Astronomical Observatories, Chinese Academy of Sciences,
Beijing 100012, P.\,R.\,China\\
$^{2}$University of the Chinese Academy of Sciences, Beijing, 100049, China, P.\,R.\,China\\
$^{3}$Instituto de Astrof\'\i sica de Canarias, E-38205 La Laguna, Tenerife, Spain\\
$^{4}$Departamento de Astrof\'\i sica, Universidad de La Laguna, E-38206 La Laguna, Tenerife, Spain\\
$^{5}$LSST, 950 North Cherry Avenue, Tucson, AZ 85719, USA \\
}

\begin{document}

\date{Accepted 2018 March 16. Received 2018 March 11; in original form 2018 February 7}

\pagerange{\pageref{firstpage}--\pageref{lastpage}} \pubyear{2018}
\maketitle
\label{firstpage}

\begin{abstract}
We present a three dimensional velocity analysis of Milky Way disk kinematics using LAMOST K giant stars and the GPS1 proper motion catalogue. We find that Galactic disk stars near the anticenter direction (in the range of Galactocentric distance between $R=8$ and 13 kpc and vertical position between $Z=-$2 and 2 \,kpc) exhibit asymmetrical motions in the Galactocentric radial, azimuthal, and vertical components. 
Radial motions are not zero, thus departing from circularity in the orbits; they increase outwards within $R\lesssim 12$ kpc, show some oscillation in the northern ($0 < Z < 2$~kpc) stars, and have north-south asymmetry in the region corresponding to a well-known nearby northern structure in the velocity field. 
There is a clear vertical gradient in azimuthal velocity, and also an asymmetry that shifts from a larger azimuthal velocity above the plane near the solar radius to faster rotation below the plane at radii of 11-12 kpc.
Stars both above and below the plane at $R\gtrsim 9$ kpc exhibit net upward vertical motions.
We discuss some possible mechanisms that might create the asymmetrical motions, such as external perturbations due to dwarf galaxy minor mergers or dark matter sub-halos, warp dynamics, internal processes due to spiral arms or the Galactic bar, and (most likely) a combination of some or all of these components.
\end{abstract}
 

\section{Introduction}
\label{introduction}
We know that there are non-axisymmetries in the Milky Way disk, 
since many velocity substructures, streams, and moving groups such as Coma Berenices, Sirius, Hyades, Pleiades, HD 1614, Arcturus were discovered. One of the most famous is the Hercules stream \citep{Denhen98, Fux01, Antoja12, Xia15}, which might be caused by the Galactic bar's outer Lindblad resonance (OLR). Other asymmetric structures can be created by internal perturbations due to the bar or spiral arms \citep {Denhen00,Fux01,Quillen05}, by external minor mergers such as the Sagittarius dwarf galaxy passing by, or by interaction with the Magellanic Clouds \citep{Gomez121, Gomez122, Minchev09, Minchev10}.

Radial asymmetrical motions have been found in RAVE data for stars within 1~kpc from the Sun. \citet{Siebert11} discovered that the mean radial velocity is larger than 10 km s$^{-1}$ from the Sun toward the Galactic centre, while it becomes smaller than $-$10 km s$^{-1}$ from the Sun toward the Galactic anti-centre, which might be caused by the two-armed spiral perturbation in which the Sun is again close to the inner ultra-harmonic 4:1 resonance \citep{Siebert12}.  \citet{Liu171} found that the younger F-type main-sequence stars have a quite different in-plane velocity field than the older sub-giant branch stars in the solar neighbourhood local spiral arm. Beyond the solar neighbourhood, \citet{Lop16} and \citet{Tian172} used red clump populations to show that the mean radial velocity is negative within R $\sim$ 9 \,kpc and positive beyond. This is likely because of the perturbation induced by the rotating bar. The zero-crossing radius of the velocity, R $\sim$ 9 \,kpc, essentially indicates the rough location of the Outer Lindblad Resonance. Non-circular streaming motions and radial streaming motions of gas have also been discovered in external galaxies \citep{Adler96, Tra08,Sellwood10}.

Vertical asymmetrical motions have also been identified in recent years. For example, the Galactic disk has been found to be oscillating in both stellar density and kinematics in the solar neighbourhood \citep{Widrow12}; the use of these ``ripples'' in the disk to identify the perturber has come to be known as Galactoseismology. \citet{Williams13} detected a rarefaction-compression
behaviour in the vertical velocity pattern at $R \lesssim$ 9 \,kpc based on red clump stars from RAVE, which they identified as a ``breathing'' mode with oscillations in $<Z^{2}>$, $<Zv>$ and $<v^{2}>$ \citep{Banik17}. This mode has odd parity in the $V_Z$ distribution and even parity in the density distribution. This is in contrast to what is known as a ``bending'' mode, with oscillations in $<Z>$ and $<v>$ \citep{Widrow14, Banik17}, and even parity in $V_Z$ with odd parity in the density distribution. With kinematics of F-type stars from the LAMOST survey, similar vertical asymmetrical substructures were also found by \citet{Carlin13}. The wave-like pattern in the mean vertical density of stars as mapped in detail with SDSS data by  \citet{Xu15} shows more stars in the north (i.e., above the Galactic plane) at distances of about 2 \,kpc (North near) from the Sun, more stars in the south (below the plane) at 4$-$6 \,kpc (South middle) from the Sun, then excess stars in the north at distances of 8$-$10 \,kpc from the Sun, and an overdensity in the south at distances of 12$-$16 \,kpc from the Sun. The residual of the density profiles after subtracting the best fit models show different oscillating patterns in almost all radii that have been probed \citep{Wang17}.

There is not yet consensus regarding the physical scenario for producing $V_Z$ and $V_R$ asymmetries. \citet{Tian15} showed that asymmetric motion may be related to the age
of stars with different dynamical relaxation times. The nearby asymmetric radial motions can be explained by the perturbation due to either the spiral arms or the bar \citep{Siebert12, Gomez13, Faure14, Monari16}. However, the asymmetric radial motions in outer disk are likely mainly contributed by the bar dynamics with a given pattern speed \citep{Grand15, Monari15, Tian172, Liu18}. The explanation must be more complex for vertical asymmetrical structure, including scenarios such as minor mergers due to the passing of the Sagittarius dwarf galaxy proposed by \citet{Gomez13} or the LMC \citep{Laporte18}, or the disk response to bombardment by merging lower-mass satellites \citep{Donghia16}. The effects of even lower-mass dark matter subhalos have also been invoked as  a possible explanation \citep{Widrow14}. 
The stellar disk has a clear warp \citep{Lop02}. According to simple analyses of vertical velocities \citep{roskar10,Lop141}, the kinematical signature of the Galactic warp's line-of-nodes is located close to the Galactic anti-center. This is also a possible explanation of the observed vertical bulk motions.

Some possible mechanisms contributing to $V_R$ are also probably influencing the azimuthal velocity. Recently, \citet{Tian172} found an intriguing U-shape along the Galactic anti-center direction, and pointed out that the reshaping of the orbits at different radius may lead to non-zero and oscillating mean in-plane velocities in the disk. \citet{Lop142} also used red clump stars to find an almost flat rotation curve with a slight fall-off, including a puzzling low average rotation speed at the outermost and most off-plane regions which might invite us to reconsider the dark matter distribution. Both of these works show that we need more data and higher precision proper motions to map the disk kinematical details.

In this work, we will further investigate radial, azimuthal and vertical velocities in
the Galactic disk. We are aspiring to figure out whether there are velocity asymmetrical motions as far away as 5 \,kpc from the solar location at low Galactic latitudes. The LAMOST spectroscopic survey \citep{Cui12, Zhao12} is used to select a larger sample of K giant stars with line-of-sight velocities and metallicities than was previously available. After these K-giants are combined with the GPS1 proper motion catalogue, which has unprecedented precision before Gaia DR2 \citep{Tian171}, we can decipher the outer disk 3-dimensional kinematical structure and asymmetries in detail. 

The paper is organised as follows. Section 2 describes our K giant star sample and coordinate transformations. Section 3 shows the 3-dimensional (3D) velocity distribution projected onto 2D maps. Then we show the results of 1D velocity distributions in Section 4. In Section 5, discussions about the asymmetrical mechanisms are included. Finally, brief conclusions are given in Section 6.

\section{Sample, distance and velocity}
\subsection{LAMOST catalogue and distance}
The Large Aperture Multi-Object Fiber Spectroscopic Telescope (LAMOST, also called the Guo Shou Jing Telescope), is a quasi$-$meridian reflecting Schmidt telescope with an effective aperture of about 4 meters. A total of 4000 fibers, capable of obtaining low resolution spectra (R $\sim$ 1800) covering the range from 380 to 900 nm simultaneously, are installed on its $5^{\circ}$ focal plane \citep{Cui12, Zhao12}. In the 5-year survey plan, LAMOST will obtain a few million stellar spectra in about 20000 square degrees in the northern sky \citep{Deng12}. The LAMOST DR3 catalogue contains 5756075 spectra; among these, the LAMOST pipeline has provided stellar astrophysical parameters (effective temperature, surface gravity, and metallicity) as well as line-of-sight velocities for about 3.2 million stars.  The sample is mainly distributed in the Galactic Anti-Center direction due to special conditions at the site \citep{Yao12}.

\begin{figure}
  \centering
  \includegraphics[width=0.48\textwidth]{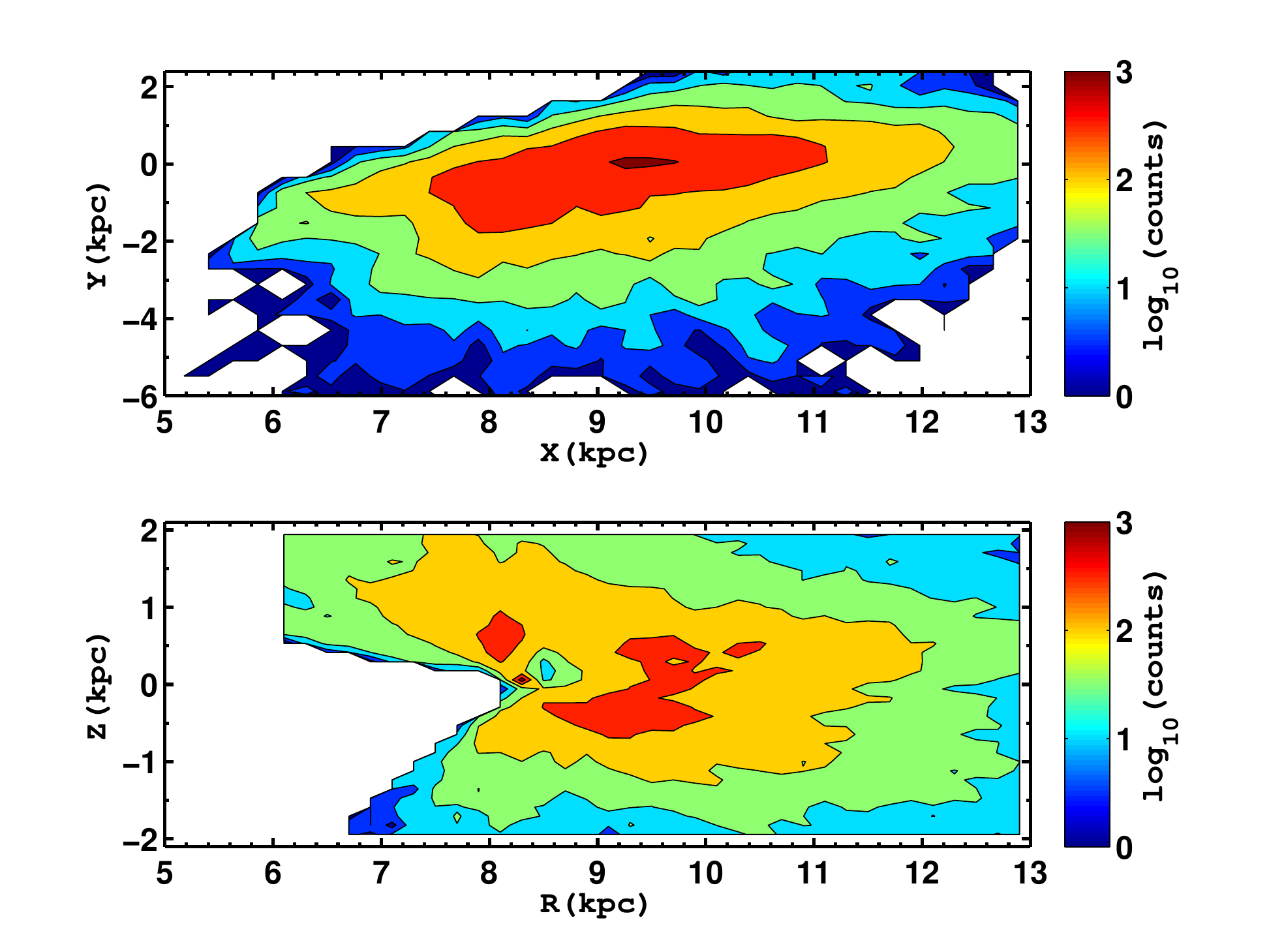}
  \caption{Spatial distribution of the 65000 K giant stars used in this work. The top panel is the $X$ and $Y$ in Galactic coordinates, and the bottom panel shows $R$ and $Z$ in cylindrical coordinates, color coded by star counts on a log scale. The majority of stars in our sample are outside the Solar radius in the direction of the Galactic Anticenter.}
  \label{XYZR}
\end{figure}

The K giant stars are selected according to the descriptions from \citet{Liu142}. Distances are estimated from a Bayesian approach \citep{Car15} with uncertainties of about 20\%. Interstellar extinction is derived using the Rayleigh-Jeans Colour Excess (RJCE) method \citep{Zasowski13} with 2MASS H band \citep{Skru06} and WISE W2-band photometry \citep{Wright10}. The difference between our derived extinction corrections and the values from \citet{Green14, Green15} is less than 0.2 mag, and the uncertainty in extinction likely contributes less than 10\% to the uncertainty in distance \citep{Liu173}. Fig.~\ref{XYZR} shows spatial distribution of the 65000 K giant stars used in this work. The top panel is looking down on the Galactic plane at Galactic $X$ and $Y$ coordinates, and shows that our sample is mainly distributed in the anti-center direction. The bottom panel of Fig.~\ref{XYZR} is the distribution in $R$ and $Z$ Galactic cylindrical coordinates, color coded by star counts on a log scale. We have adopted a Galactic coordinate system with $X$ increasing  outward from the Galactic centre, $Y$ in the direction of rotation, and $Z$ positive towards the North Galactic Pole (NGP) \citep{Williams13}.

\subsection{Proper motion and velocity transformation}
The proper motion catalog known as GPS1 (Gaia$-$PS1$-$SDSS) \citep{Tian171} has a precision of $\sim$ 1.5$-$2 mas yr$^{-1}$ with systematic error 0.3 mas yr$^{-1}$.
For this work, we match our LAMOST-selected K giants to GPS1, then calculate the 3D Galactocentric cylindrical coordinates for the K giant stars by adopting a location of the Sun of $R_{\odot}$  = 8.34 \,kpc \citep{Reid14} and $Z_{\odot}$ = 27 \,pc \citep{Chen01}. The heliocentric rectangular components of the Galactic space velocity $U, V$, and $W$ are determined by the right-handed coordinate system based on \citet{Johnson87}, with $U$ positive towards the Galactic centre, $V$ positive in the direction of Galactic rotation and $W$ positive towards the north Galactic pole. Cylindrical velocities $V_R$, $V_\theta$, and $V_Z$ are defined as positive with increasing $R$, $\theta$, and $Z$, with the latter towards the NGP. For the solar motion we use \citet{Tian15} value, [$U_{\odot}$ $V_{\odot}$ $W_{\odot}$] = [9.58, 10.52, 7.01] km s$^{-1} $. The circular speed of the LSR is adopted as 238 km s$^{-1} $ \citep{Schonrich12}. We also correct the radial velocities of the samples by adding 4.4 km s$^{-1}$ in this work which is smaller than \citet{Tian15} 5.7 km s$^{-1}$ due to the updating of the software of the data process compared with APOGEE data \citep{Blanton17}.

\begin{figure}
  \centering
  \includegraphics[width=0.48\textwidth]{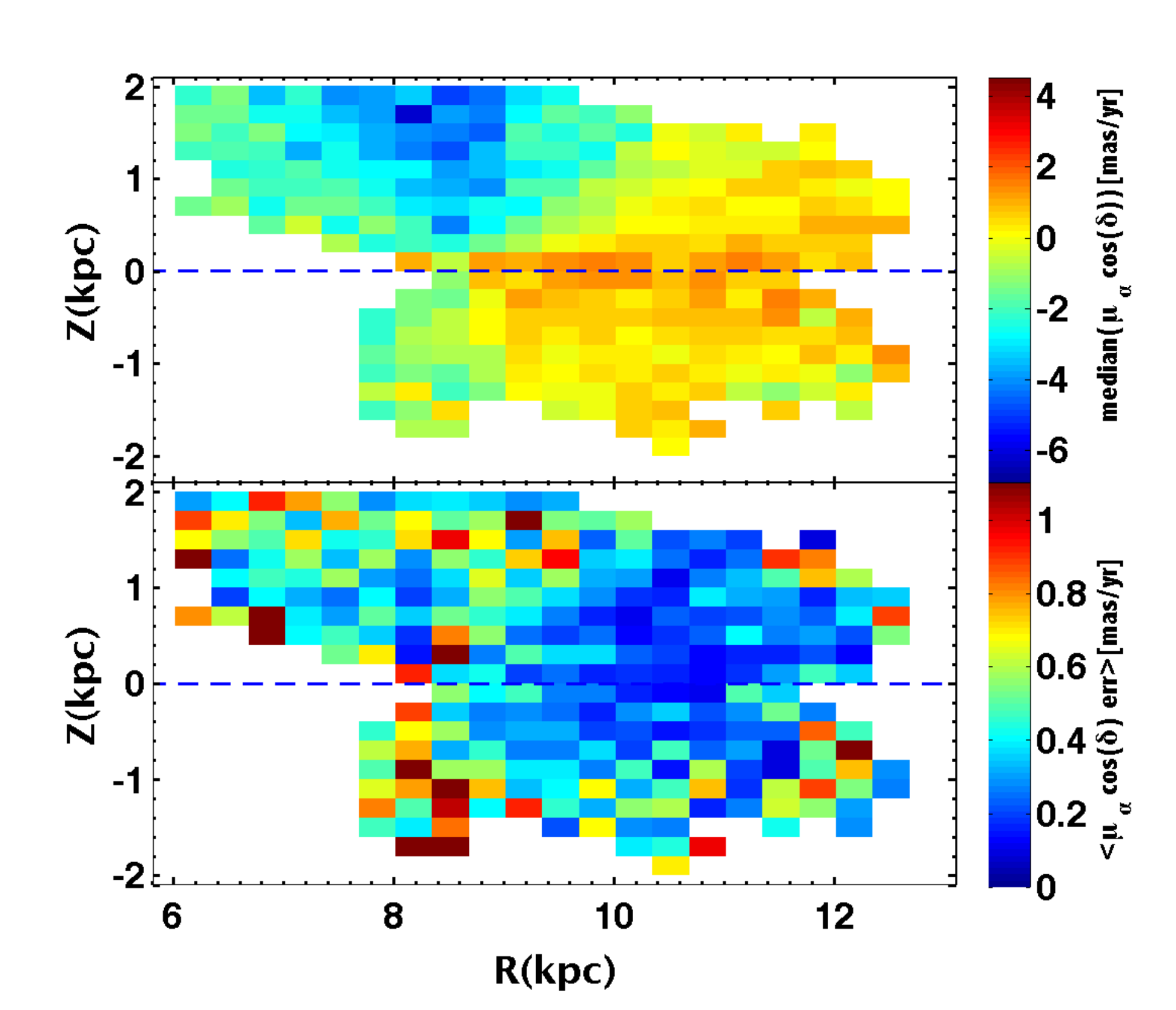}
  \caption{Behavior of proper motion measurements for K giant stars in the range of $Z=[-2, 2]$ \,kpc. The color-coded map shows the distribution of the median (top) and bootstrap error (bottom) for the right ascension proper motion components on the $R, Z$ plane. The median number of stars per pixel is larger than 50.}
  \label{RZPMRA}
\end{figure}

\begin{figure}
  \centering
  \includegraphics[width=0.48\textwidth]{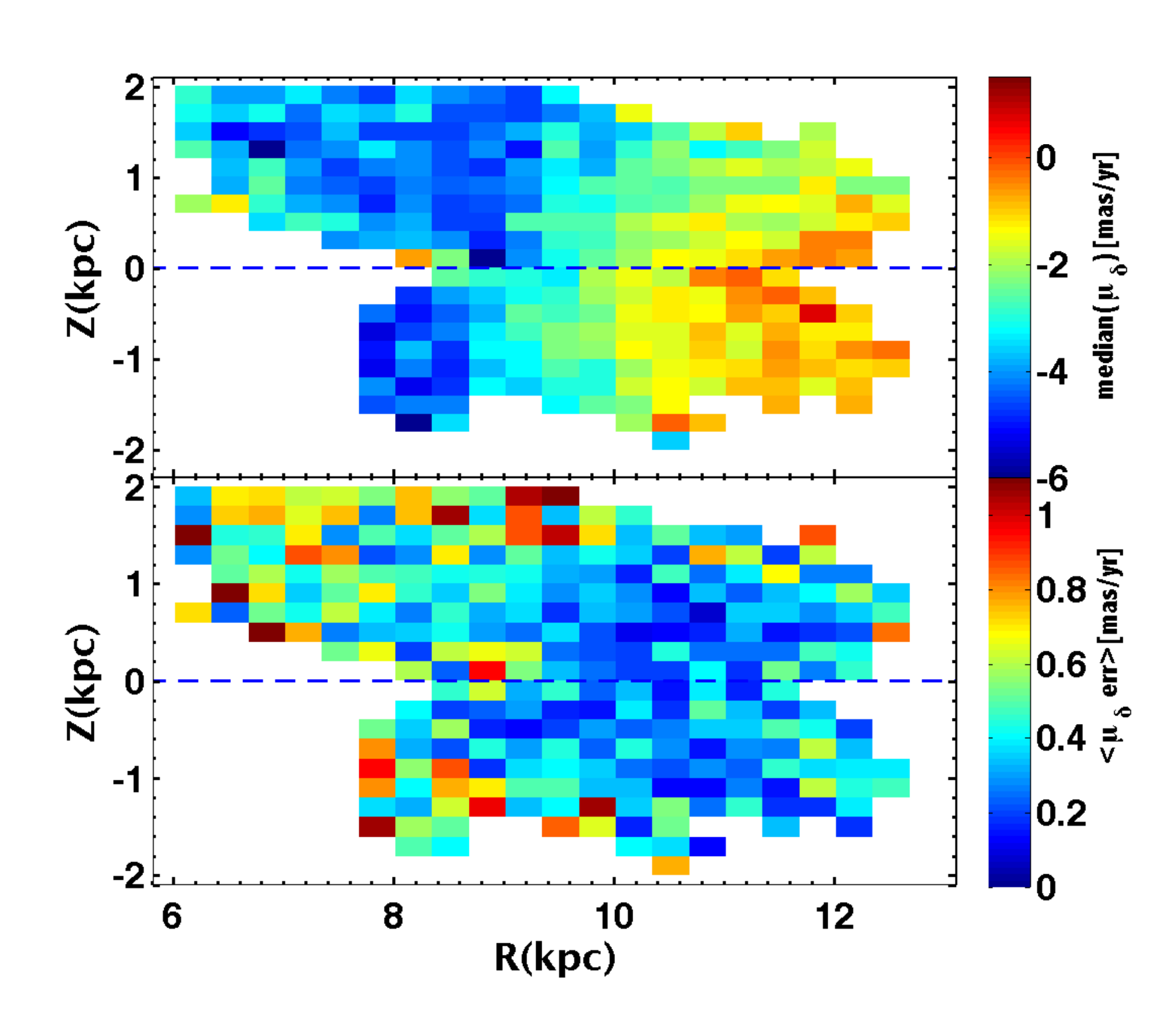}
  \caption{Similar to Fig.~\ref{RZPMRA}, but for proper motion in the declination direction.}
  \label{RZPMDEC}
\end{figure}

The behavior of proper motion measurements for K giant stars within the range of $Z = [-2, 2]$ \,kpc is shown in Figures~\ref{RZPMRA} and \ref{RZPMDEC}, with each figure showing a color-coded map of the median proper motion (top panel) and its associated bootstrap error (bottom) for the right ascension (Fig.~\ref{RZPMRA}) and declination (Fig.~\ref{RZPMDEC}) proper motion components in the $R$ (radial distance from the Galactic centre), $Z$ (vertical distance from the midplane) plane. We only use the bins of which the median number of stars per pixel is greater than 50. The choice of bin size is 0.33 \,kpc for $R$ and 0.2 \,kpc for $Z$. Bootstrap errors are determined by resampling (with replacement) 100 times for each bin, and the uncertainties of the estimates are determined using 15\% and 85\% percentiles of the bootstrap samples. For both components, most of the bootstrap errors are less than 1.2 mas yr$^{-1}$. The GPS1 catalogue systematic errors are nearly an order of magnitude better than those in PPMXL \citep{Roeser10} and UCAC4 \citep{Zacharias10}, and their random errors are roughly a four-fold improvement relative to PPMXL and UCAC4, making this the ideal catalogue for our purposes.

For our sample during this paper, the stars located inside $|Z|$ $<$ 2 kpc and 6 $<$ R $<$ 13 \,kpc are selected to map the lower disk kinematics in a region in which we have small random and systematic errors. The stars with LAMOST spectroscopic SNR $<$ 20 are not included, and we also exclude stars with [Fe/H] $<$ $-$1.0 \,dex, so we can focus on Galactic disk populations. We also set some criteria in velocity to remove fast-moving halo stars: $V_R$=[-150, 150] km s$^{-1}$, $V_{\theta}$=[-50, 350] km s$^{-1}$, and $V_Z$=[-150, 150] km s$^{-1}$; this sacrifices a small fraction of stars, yielding a final sample of around 65000 disk K-giant stars.

\section{2$-$D velocity asymmetric structure}   
\begin{figure}
  \centering
  \includegraphics[width=0.48\textwidth]{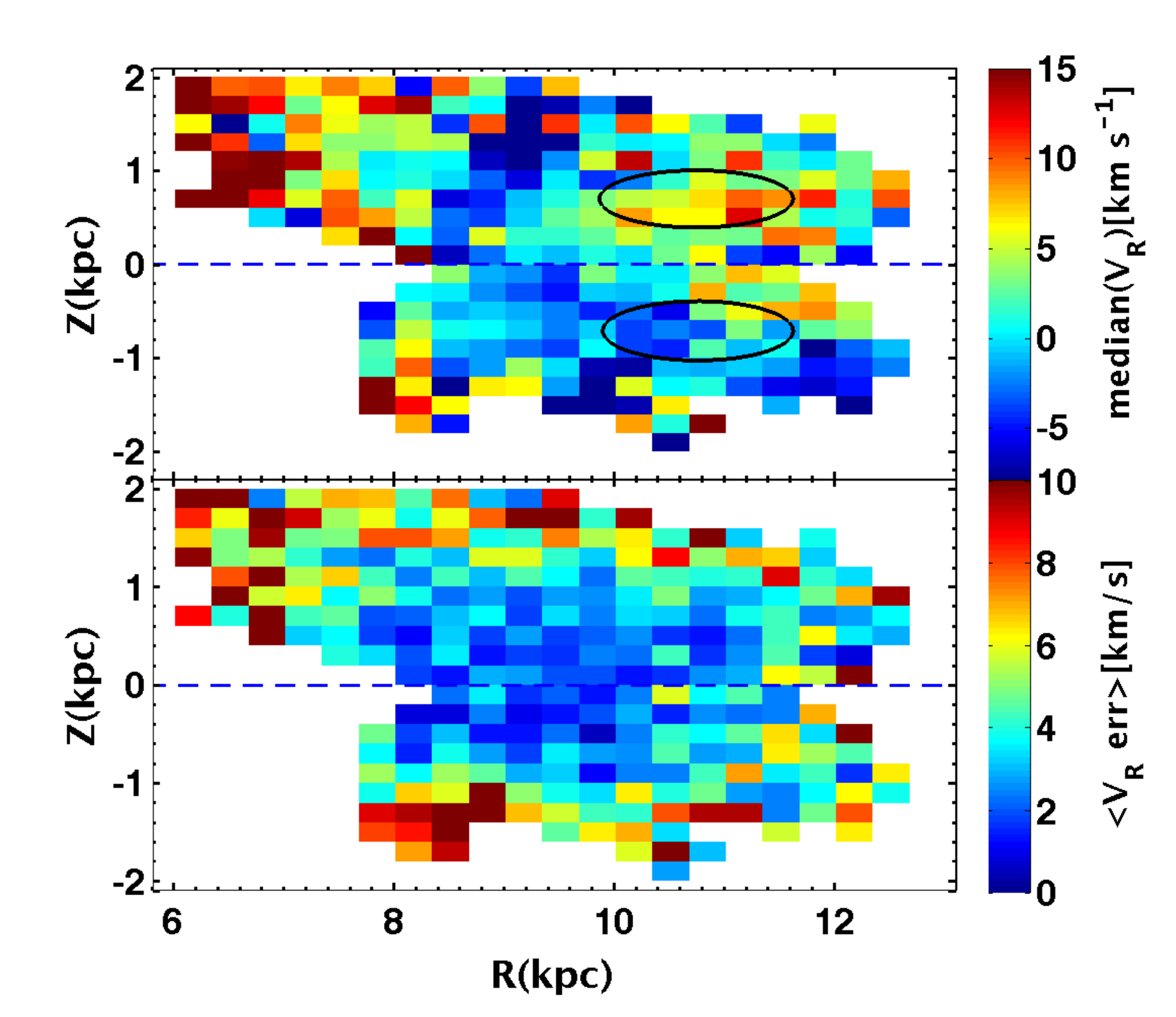}
  \caption{Average radial Galactocentric velocity in spatial bins; the upper panel shows the median $V_R$ value in each bin, and the lower panel their corresponding bootstrap errors. In this figure, the line-of-sight velocity is from LAMOST and the proper motion is from GPS1. Notice that there is an oscillating structure with radius at $0 < Z < 2$~kpc, and there is a north-south asymmetry at $R \sim$ 10-11 \,kpc, $Z \sim$ 0.5 \,kpc. The two ellipses highlight the regions where we see a north/south asymmetry.}
  \label{RZVR}
\end{figure}

In this section, we focus on describing 2D asymmetrical motions in the Galactic disk. The top panel of     
Fig.~\ref{RZVR} shows the variation of Galactocentric radial velocity $V_R$ in the $R, Z$ plane. The errors in the bottom panel are computed by a bootstrap method. The error is mainly contributed by the distance uncertainty, proper motion random errors and  statistical errors. The most prominent feature displayed in the upper panel of Fig.~\ref{RZVR} is that there is a large oscillating structure on the northern ($0 < Z < 2$ \,kpc) side. Inside the solar radius (from $6 < R \lesssim 8$~kpc), $V_R$ has a positive (mean) value, then it becomes negative from $R \sim$ 8.4 to 9.8 \,kpc (on average), then it becomes positive when $R \gtrsim$ 10 \,kpc. On the southern side, most of the bins are moving inward with negative value, mixed with a few positive bins. It is intriguing that there seems to be a negative-velocity (blue colors in Fig.~\ref{RZVR}) C-shaped feature extending from $(R,Z)$ = (9, 1) to (8, 0) to (12, -1) \,kpc, suggesting a connection between velocity features above and below the plane. We also note that there is a north-south asymmetry at $R \sim$ 10-11 \,kpc, $Z \sim$ 0.5 \,kpc (denoted by the two ellipses in the figure), corresponding to mean positive values in the north and negative $V_{\rm R}$ in the south. The average $V_R$ error in most regions is around 2 km s$^{-1}$.

\begin{figure}
  \centering
  \includegraphics[width=0.48\textwidth]{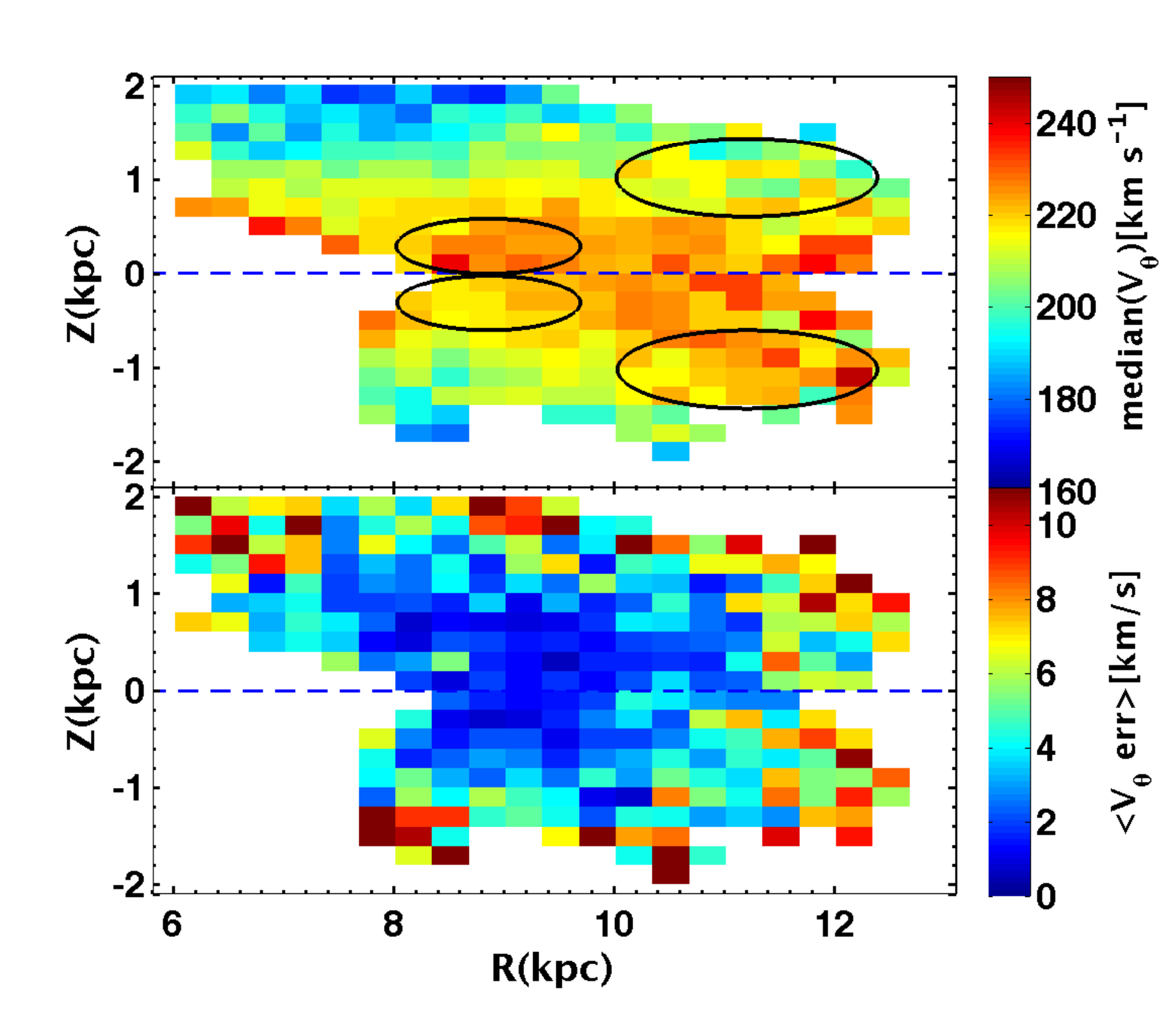}
  \caption{Similar to Fig.~\ref{RZVR}, but for azimuthal velocity. There is a clear gradient with $Z$ representing the transition from thin to thick disk. The four ellipses highlight regions exhibiting north $-$ south asymmetries in their $V_\theta$ velocities.}
  \label{RZVPHI}
\end{figure}

The distribution of $V_{\theta}$ (Figure~\ref{RZVPHI}) shows different trends from $V_R$, exhibiting a clear gradient with $|Z|$ on both sides of the disk that represents the transition from thin to thick disk. The in-plane kinematical features of $V_{\theta}$ in the Galactic outer disk shown in \citet{Tian172} are not seen in this panel. There are clear asymmetrical rotational motions: when $R \geq$ 10 \,kpc and $|Z| \geq$ 0.5 kpc, stars in the south are rotating faster than those at symmetrical locations in the north by $\sim$ 10-40 km s$^{-1}$. At $R  \approx $ 8 $-$ 9\, kpc and $Z \lesssim$ 0.5 \,kpc, there is a similar trend but reversed; here, the northern stars are faster than those in the south, but the feature is not as clear as in the outer region. 
We note that the asymmetric motions we are discussing in relation to Figure~\ref{RZVPHI} are in addition to the well-known ``asymmetric drift'' that is responsible for the vertical gradient in $V_\theta$. The asymmetrical motions we refer to are differences between the mean velocities at symmetrically located positions on either side of the Galactic plane.
We show ellipses in Figure~\ref{RZVPHI} to help the reader identify the features we are discussing. We will discuss these further in the next section (see, e.g., Fig.~\ref{1DVPHI_test1}).

\begin{figure}
  \centering
  \includegraphics[width=0.48\textwidth]{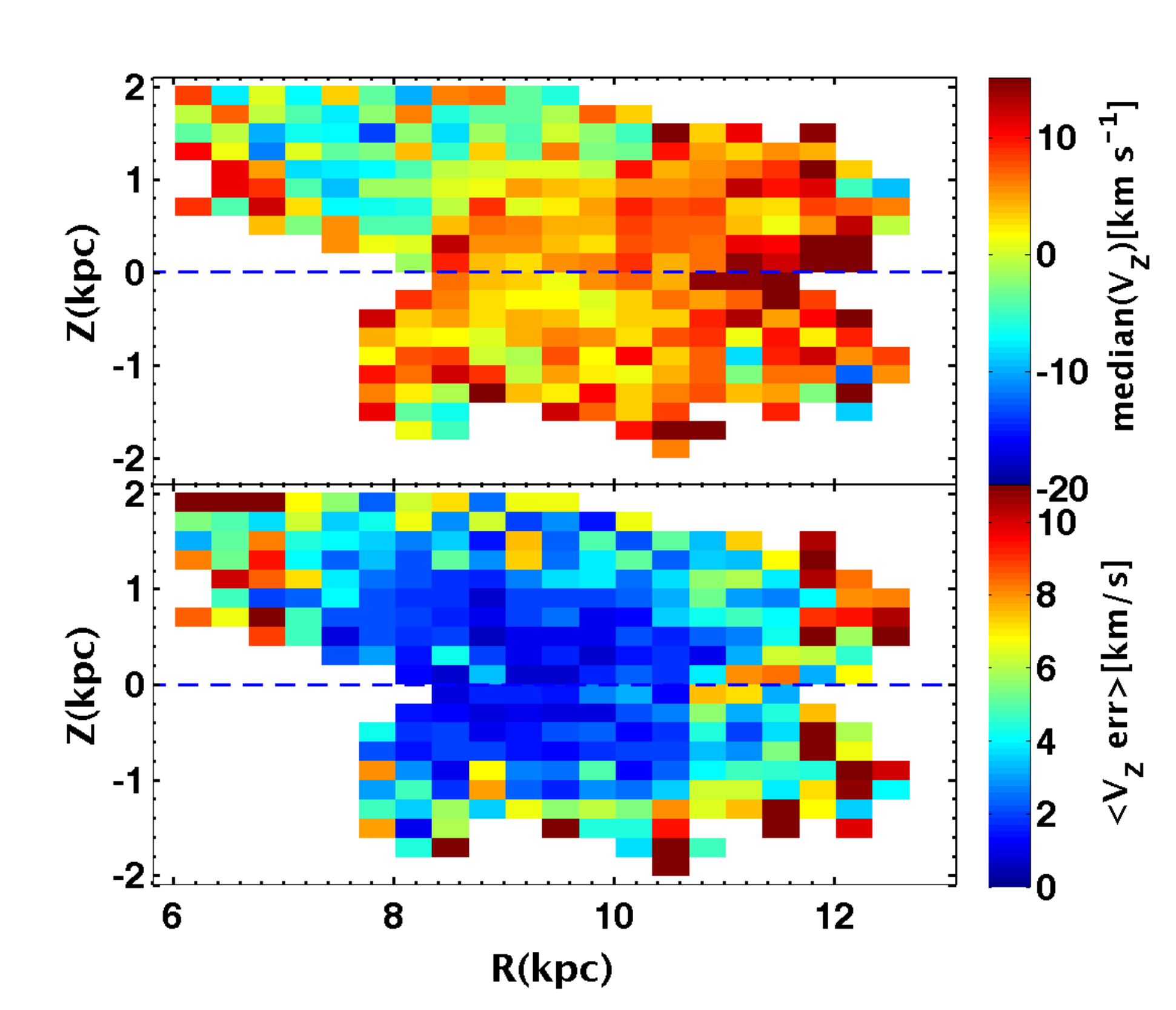}
  \caption{Similar to Fig.~\ref{RZVR}, but for vertical velocity, $V_Z$. Notice that there is compression motion inside the solar circle, and significant bulk motions at all radii and heights outside the Sun's radius, extending until $R \sim$ 13 \,kpc.}
  \label{RZVZ}
\end{figure}

Figure~\ref{RZVZ} shows the average $V_{Z}$ with position in the $R, Z$ plane. There is clear evidence of a compression motion inside the solar radius: above the midplane, stars are moving downward, and below the plane, stars are moving upward. This is very similar to substructure found by \citet{Carlin13}. Some works describe this as the ``breathing'' mode \citep{Widrow14, Carrillo17}, but in this work we will simply call it asymmetrical motion. Outside the solar location, we can see in the figure that stars are on average moving upward at almost all radii for $|Z|\le $2 \,kpc, which is partly different from \citet{Carlin13}. This illustrates what is often called a ``bending'' mode. This is comparable to the vertical bulk motion maps in Fig. 4 of \citet{Lop141}, but the accuracy of their results is low due to the large PPMXL proper motion errors, and they mentioned they only reported a tentative detection of vertical motion with low significance. We discuss some possible mechanisms to create the bending modes observed in Figure~\ref{RZVZ} in the discussion.

\section{1$-$D velocity asymmetric structure}
\subsection{Velocity profiles in the vertical direction}
In this section, we continue to describe the trends in median velocity as a function of $(R, Z)$ position in the Galaxy for K giant stars with GPS1 catalogue proper motions. Error bars give the measured bootstrap errors. We set a bin size of 0.2 \,kpc, and keep only bins with star counts larger than 50. When we have low density of stars, in order to keep at least 50 stars per bin, the $R$ bin is set to 2 \,kpc, which is relatively larger than other works \citep{Liu18, Tian172}. It will not affect us when discussing low latitude outer disk asymmetrical kinematics. In the figures of this section, we have used stars with [Fe/H] $>$ $-$0.4 dex and we also tested to consider other solar motions \citep{Schonrich10}. The default value we set for the circular velocity of the local standard of rest (LSR) is 220 km s$^{-1}$, with a solar radius of 8 \,kpc, but we have used other values and we have seen that the overall trends are stable in this work by changing those parameters.

\begin{figure}
  \centering
  \includegraphics[width=0.48\textwidth]{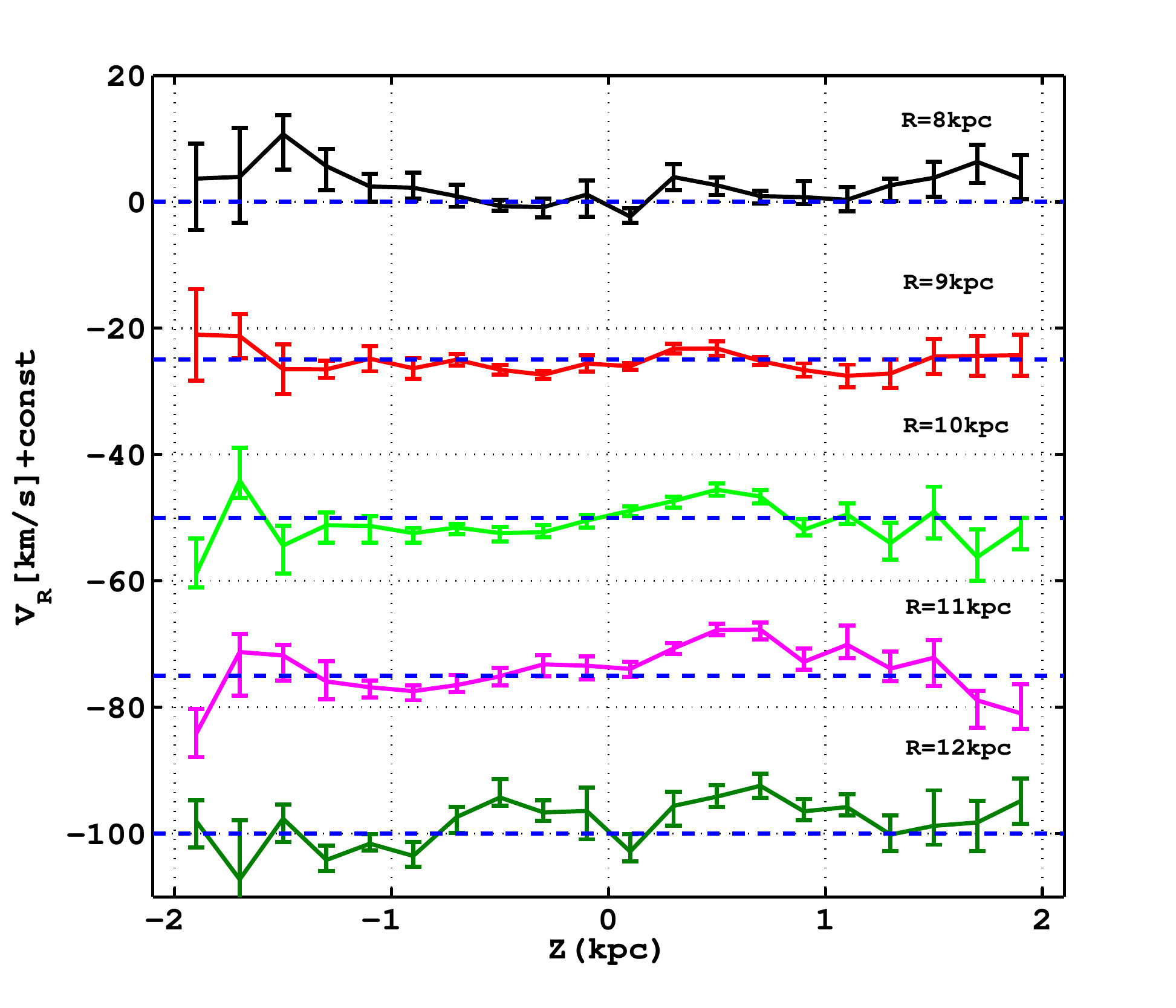}
  \caption{The trends in median velocity as a function of ($R, Z$) position in the Galaxy for K giant stars with GPS1 catalogue proper motions. Error bars give the measured bootstrap errors. The bin size is 0.2 \,kpc with star counts required to be larger than 50 per bin. Different colors represent median values of $V_R$ as a function of $Z$ for different $R$ ranges, with [Fe/H] $>$ $-$1.0 dex. Notice that at $R \sim$ 10-11 \,kpc, $Z \sim$ 0.5 \,kpc, there is an asymmetrical velocity substructure, with inward motions below the plane, and outward $V_R$ above the plane. Velocities for each $R$ slice have been shifted by a constant value; the dashed line represents $V_R = 0$~km~s$^{-1}$.}
  \label{1DVR}
\end{figure}

\begin{figure}
  \centering
  \includegraphics[width=0.46\textwidth]{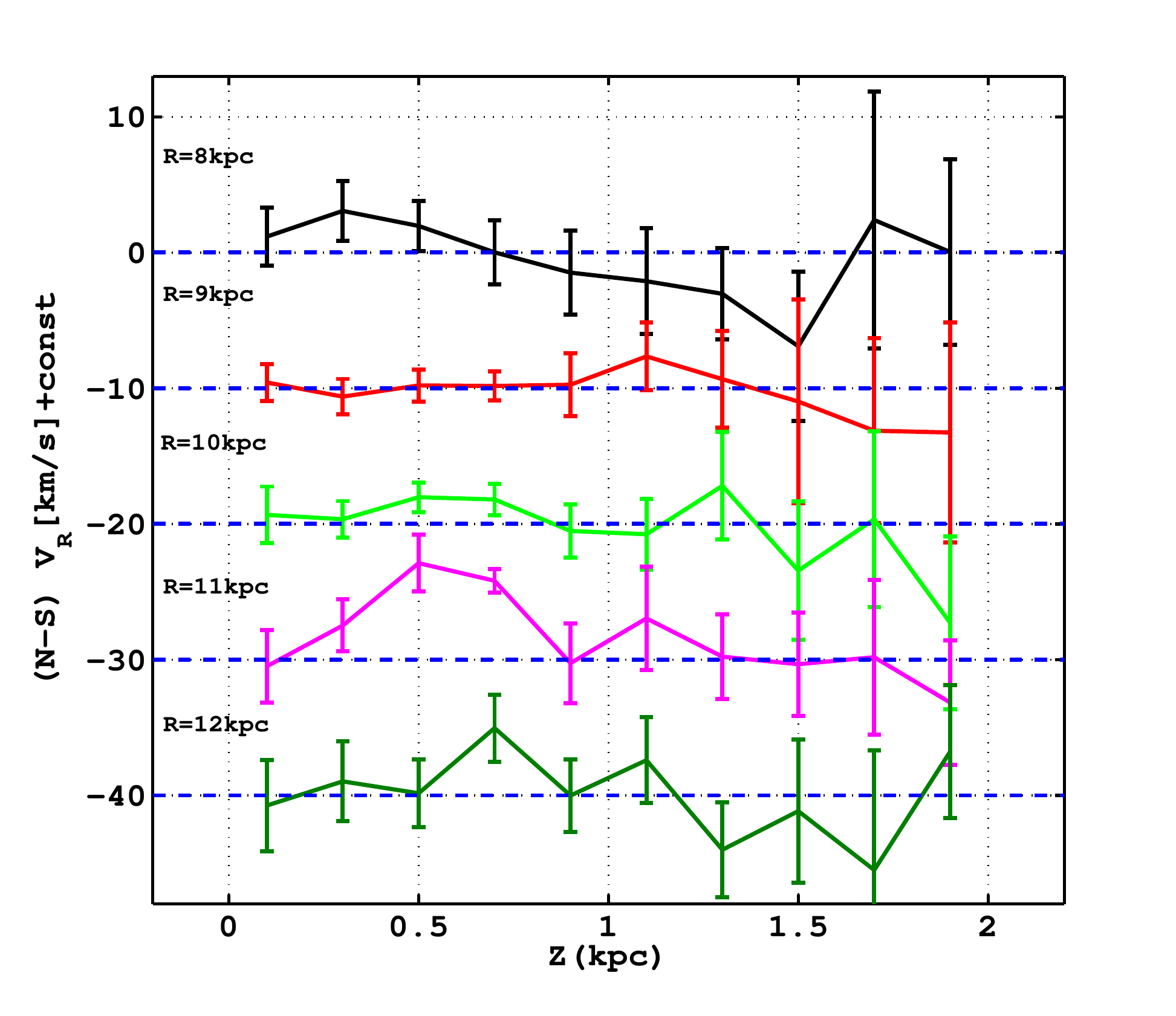}
  \caption{Difference between median $V_{R}$ velocities in North-South symmetric regions as a function of $|Z|$ in different distance slices. At $R\approx$ 11 \,kpc, $Z\approx$ 0.5 kpc there is a clear velocity substructure corresponding to the North-near structure of \citet{Xu15}.}
  \label{1DVR_test}
\end{figure}

The median $V_R$ vs. $Z$ at $R=8-12$ \,kpc is shown in Fig.~\ref{1DVR}. Different colors represent median values of $V_R$ as a function of $Z$ for different $R$ ranges. We can see that at $R \sim$ 10-11 \,kpc, $Z \sim$ 0.5 \,kpc, there is a north-south asymmetrical velocity substructure which is consistent with the Fig.~\ref{RZVR} substructure mentioned before, and there is an almost symmetric outward motion at $R=12$ \,kpc. For other locations, there is an almost flat $V_R$ curve and no significant asymmetric motion. In order to quantitatively describe this behaviour, in Fig.~\ref{1DVR_test} we show the North - South (N - S) difference in median $V_R$ with $|Z|$ at different $R$. There is clear substructure around $R \sim$ 11 \,kpc, $Z \sim$ 0.5 \,kpc; at 10 or 12 \,kpc, there is a similar trend but weaker.

\begin{figure}
  \centering
  \includegraphics[width=0.48\textwidth]{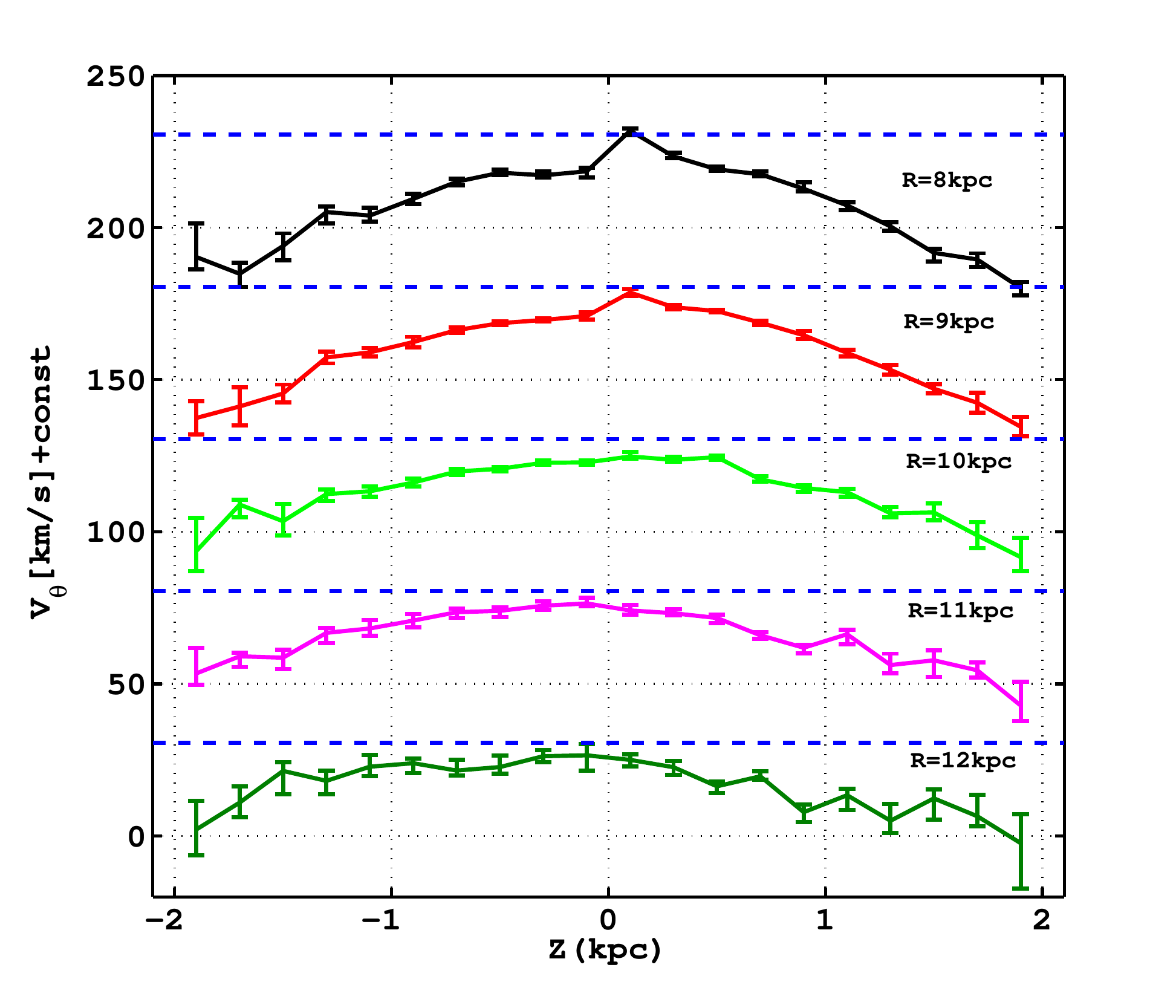}
  \caption{Similar to Fig.~\ref{1DVR}, but for azimuthal velocity. At distances larger than 10 \,kpc, there is an asymmetry when comparing regions at similar heights above and below the plane. These asymmetries vary in magnitude and sign as a function of $R$. }
  \label{1DVPHI}
\end{figure}

\begin{figure}
  \centering
  \includegraphics[width=0.46\textwidth]{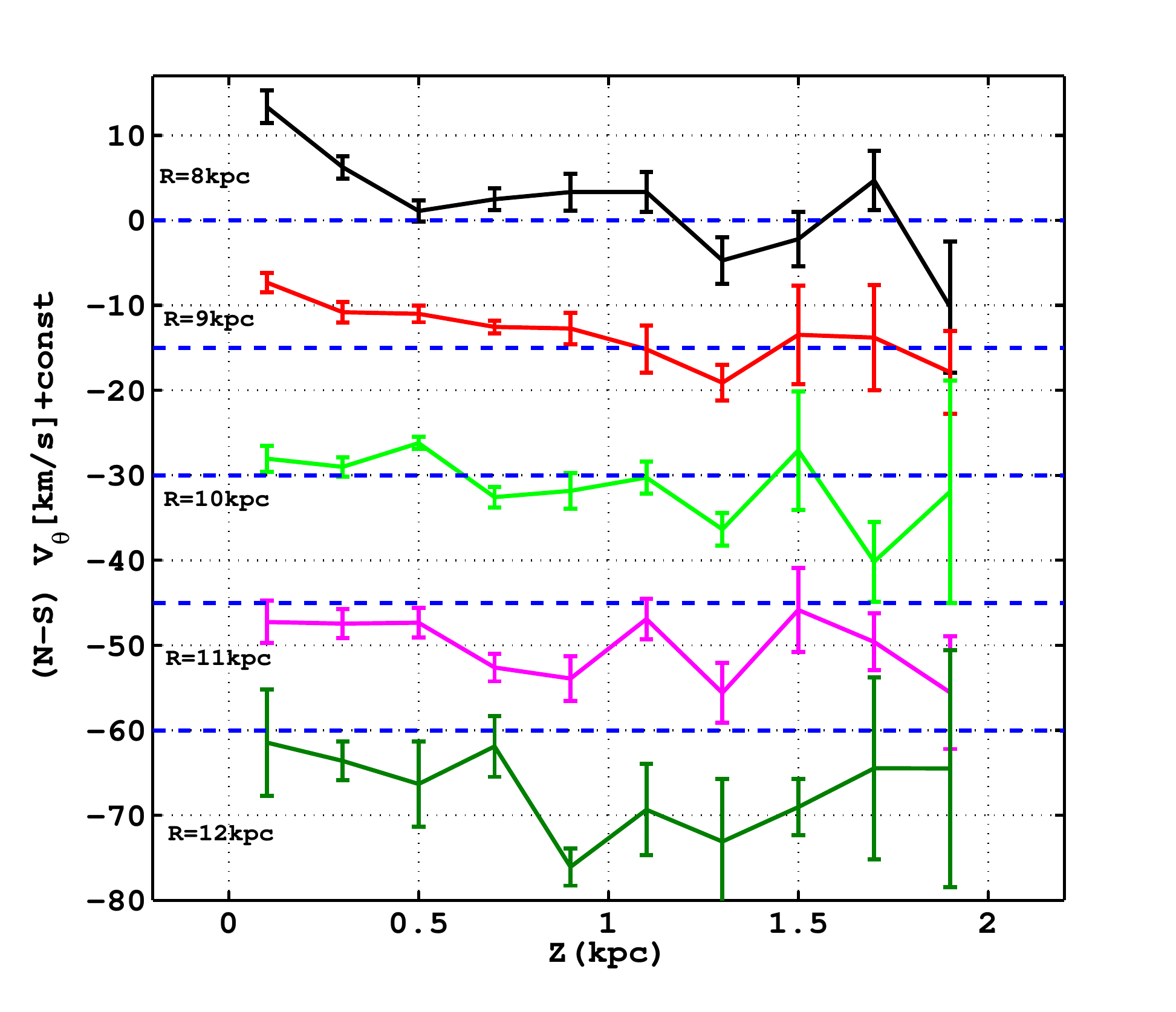}
  \caption{Similar to Fig.~\ref{1DVR_test}, but for azimuthal velocity. All stars in the south are rotating faster than those in the north for $R=11$ and 12 \,kpc (corresponding to negative values of $V_{\theta, north} - V_{\theta, south}$), although some points have large error bars.}
  \label{1DVPHI_test1}
\end{figure}

The median $V_{\theta}$ vs. $Z$ at $R=8-12$ \,kpc is shown in Fig.~\ref{1DVPHI}. The curves at $R \sim$ 8-9 \,kpc are almost symmetrical about $Z=0$~kpc. These transition to an asymmetrical distribution for distances larger than 10 \,kpc, where the median $V_\theta$ in the south is typically larger than that in the north (on average) at a given radius. As we did for $V_R$, we compare the median $V_\theta$ for northern stars to southern stars in Fig.~\ref{1DVPHI_test1}. For almost all $|Z|$ bins the north-south velocity residuals are negative at $R=11, 12$~kpc; this implies that the stars in the south are rotating faster than those in the north. At R $ \approx $ 8 $-$ 9 \,kpc and $|Z| \lesssim$ 0.5 \,kpc, there is a similar trend but reversed pattern showing northern stars are rotating faster than those in the south. We speculate it is likely a real effect in the discussion.

\begin{figure}
  \centering
  \includegraphics[width=0.48\textwidth]{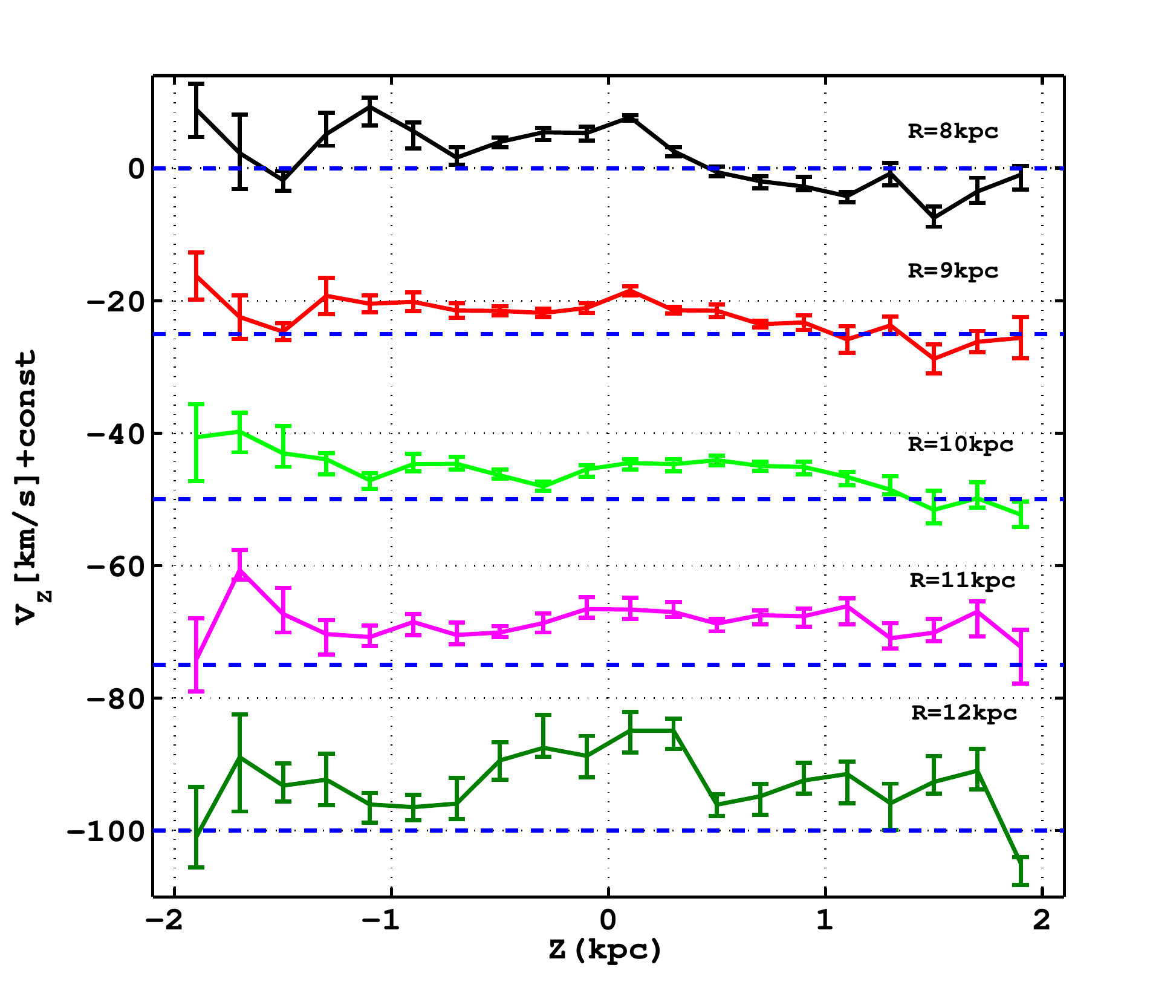}
  \caption{Similar to Fig.~\ref{1DVR}, but for vertical velocity. At $R=8$ \,kpc there is a clear oscillation from positive velocities below the plane to negative $V_Z$ above the plane. For $R=9$ \,kpc, the trend is similar to that at 8 \,kpc, but the pattern is weaker. From 10 \,kpc outward, there are significant upward bulk motions in the range $-$2 \,kpc $<$ Z $<$ 2 \,kpc.}
  \label{1DVZ}
\end{figure}

Median $V_{Z}$ vs. $Z$ at $R=8-12$ \,kpc is shown in Fig.~\ref{1DVZ}. The median $V_Z$ velocity for $R=8$ \,kpc oscillates from a positive value of $\sim$ 8 km s$^{-1}$ to negative ($\sim -5$ km s$^{-1}$) between $Z = -2$~\,kpc and $Z = 2$~\,kpc. The behaviour is similar for $R=9$ \,kpc, but the trend at 9 \,kpc is weaker. For $R >$ 9 \,kpc, there are significant upward motions from $V_Z \sim $ 3 km s$^{-1}$ to 15 km s$^{-1}$, especially between $-1 < Z < 1$ \,kpc. The trends at $R = 8$ \,kpc are similar to what is typically known as a breathing mode, while those at $R>9$~kpc may represent a bending mode. In this work, we will simply refer to these as vertical asymmetrical motions in general. Some possible mechanisms to excite these motions are discussed in the next section. 

\begin{figure*}
  \centering
  \includegraphics[width=0.98\textwidth]{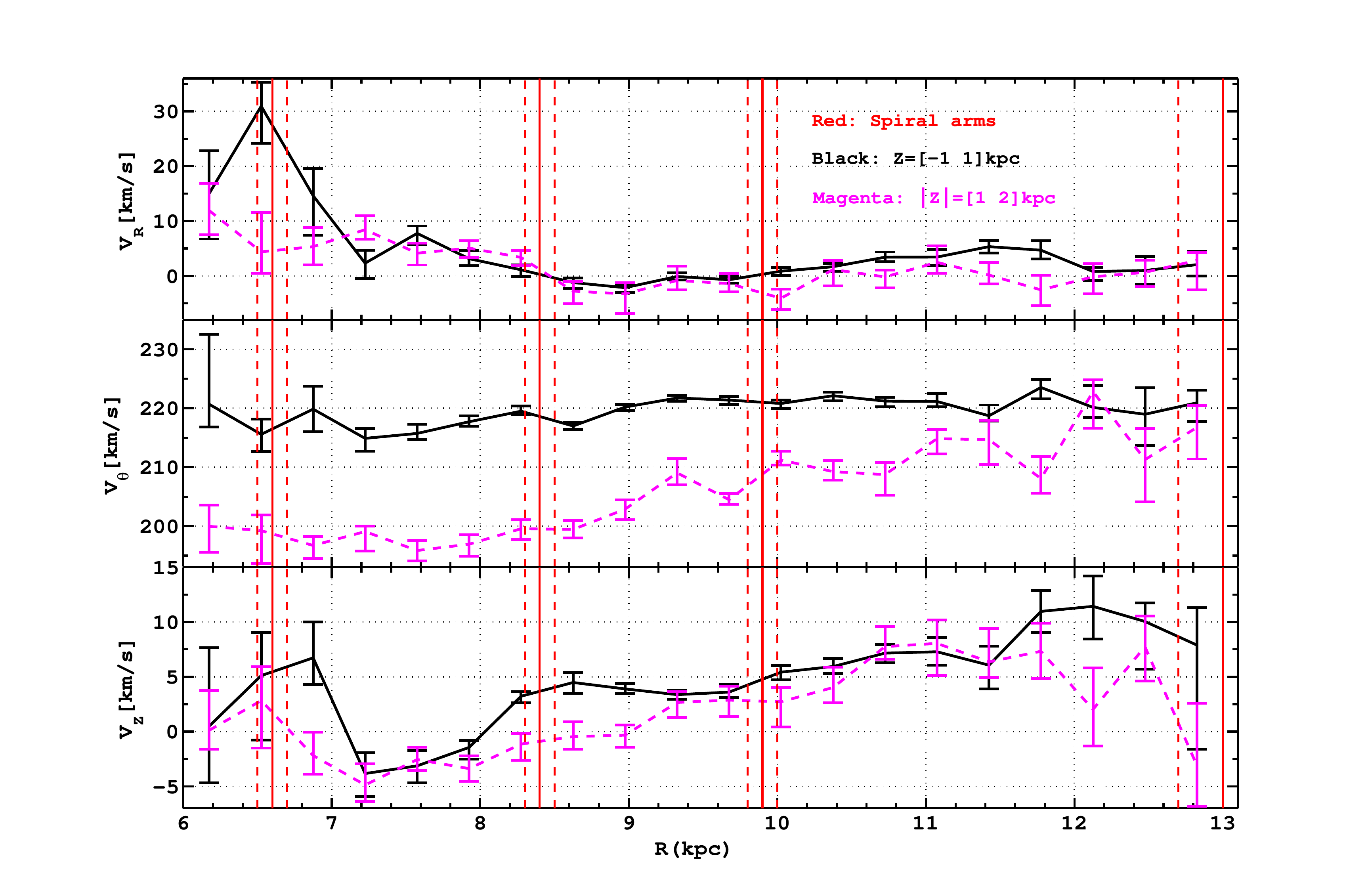}
  \caption{The top panel shows the variation of median $V_R$ for K giant stars with [Fe/H] $>$ $-$1.0 dex in the range of $Z$=[$-$2 $+$2] \,kpc and $R$=[6 13] \,kpc. The black solid curve with error bars represents the stars in the range of $Z$=[-1, 1] kpc, and the magenta dashed curve represents stars in the range $Z$=[-2, -1] \,kpc and $Z$=[1, 2] \,kpc. The middle panel shows the variation of median $V_{\theta}$, which ranges from 198 km s$ ^{-1} $ increasing to 220 km s$ ^{-1} $. The stars near the midplane exhibit little change in median $V_\theta$ with radius, while stars further from the plane (the magenta curve) show a clear variation with radius, along with an overall difference indicative of the rotational velocity gradient with height at all radii. The bottom panel shows the vertical velocity $V_{Z}$ along $R$. There is a clear trend from negative trend to positive trend with upward bulk motions at all radii $R \gtrsim 9$~kpc. The red vertical solid and dashed lines in all panels mark the locations of spiral arms \citep{Reid14}.}
  \label{1DRVRVPHIVZ}
\end{figure*}

\subsection{Velocity profiles in the radial direction}
The radial profiles of the binned median $V_R$, $V_{\theta}$, and $V_{Z}$ for stars with [Fe/H] $>$ $-$1.0 dex in the range of $Z$ = [$-$2, $+$2] \,kpc and $R$=[6, 13] \,kpc are shown in Fig.~\ref{1DRVRVPHIVZ}. In order to see more details and avoid the influence of moving groups or streams on our discussion, we divide our samples into the lower region stars ($Z$ = [-1, 1] \,kpc) and higher region stars ($|Z|$ = [1, 2] \,kpc).

The top panel of Fig.~\ref{1DRVRVPHIVZ} shows the variation of median $V_R$ for K giant stars. The red vertical solid and dashed lines in all panels mark the locations of spiral arms \citep{Reid14}. The solid black line with error bars represents stars in the range of $Z$ = [-1, 1] \,kpc, and the magenta dashed line and error bars represent stars in the range $Z$ = [-2, -1] \,kpc and $Z$ = [1, 2] \,kpc. The top panel shows that lower region stars have larger average $V_R$ than the higher region stars in the ranges of R $\sim$ 6 to 7.2 \,kpc and 10 to 12 \,kpc. At other radii, there are only small differences between $V_R$ of the lower/higher samples. The middle panel displays the variation of median $V_{\theta}$ with $R$. The value ranges from $\sim$198 km s$^{-1} $ to 220 km s$ ^{-1}$ for the higher region stars, while the lower region stars have small variations from an average of $\sim220$ km s$ ^{-1}$.  There are clear differences between the lower-$Z$ stars and higher stars which reflect the rotational velocity gradient with height from the midplane. The bottom panel shows the vertical velocity $V_{Z}$ along $R$. There is a clear trend with radius from negative (downward) to positive (upward) bulk motions. Furthermore, the vertical velocities of the lower region stars are larger than those in the higher region at all radii except $R \sim 10.8$ to 11.4 \,kpc. 
Some possible scenarios to explain the trends noted here are discussed in the next section. We note that we only concentrate on stars at $R<13$ \,kpc. At distances larger than $\sim13$ \,kpc, we do not have enough stars in our sample, and the errors become too large to see small-amplitude velocity trends.

\section{Discussions}
\subsection{Possible implications and comparisons}
Radial velocity oscillations might be produced by the spiral structures that are always spatially correlated with spiral arms \citep{Siebert12, Faure14, Grand15}. However, the top panel of Fig.~\ref{1DRVRVPHIVZ} does not show such a correlation with any spiral arm in the outer disk. We speculate that the spiral arms are not the main contributor to, or at least have only a small effect on, the observed asymmetrical velocity structure. The overall trends we see are similar to those found by  \citet{Tian172} among old red clump stars. Minor mergers (i.e., accreting dwarf galaxies such as the Sgr dwarf) may raise vertical waves \citep{Gomez13}, but these perturbations may not intensively affect the in-plane velocity \citep{Tian172}. Therefore, we speculate that the radial oscillation we see may be mainly induced by bar dynamics, the Galactic dark matter halo, a disk warp, or other mechanisms such that the mean orbit of the disk stars is intrinsically elliptical or there is a net secular expansion of the disk \citep{Lop16}. From the top panel in Fig.~\ref{1DRVRVPHIVZ}, we can also see that these possible effects have larger dynamical effects on the lower disk than stars further from the midplane.

Notice that there is clear asymmetrical structure in the Fig.~\ref{1DVR} around $Z\approx$ 0.5 \,kpc and $R\approx$ 10$-$11 \,kpc. We have found that this corresponds to the \citet{Wang17} overdensity, which is called the north near structure in \citet{Xu15}. The location is around 2 \,kpc from the Sun. So we find both a density and velocity signature of the north near structure simultaneously in this series of works.

\begin{figure}
  \centering
  \includegraphics[width=0.48\textwidth]{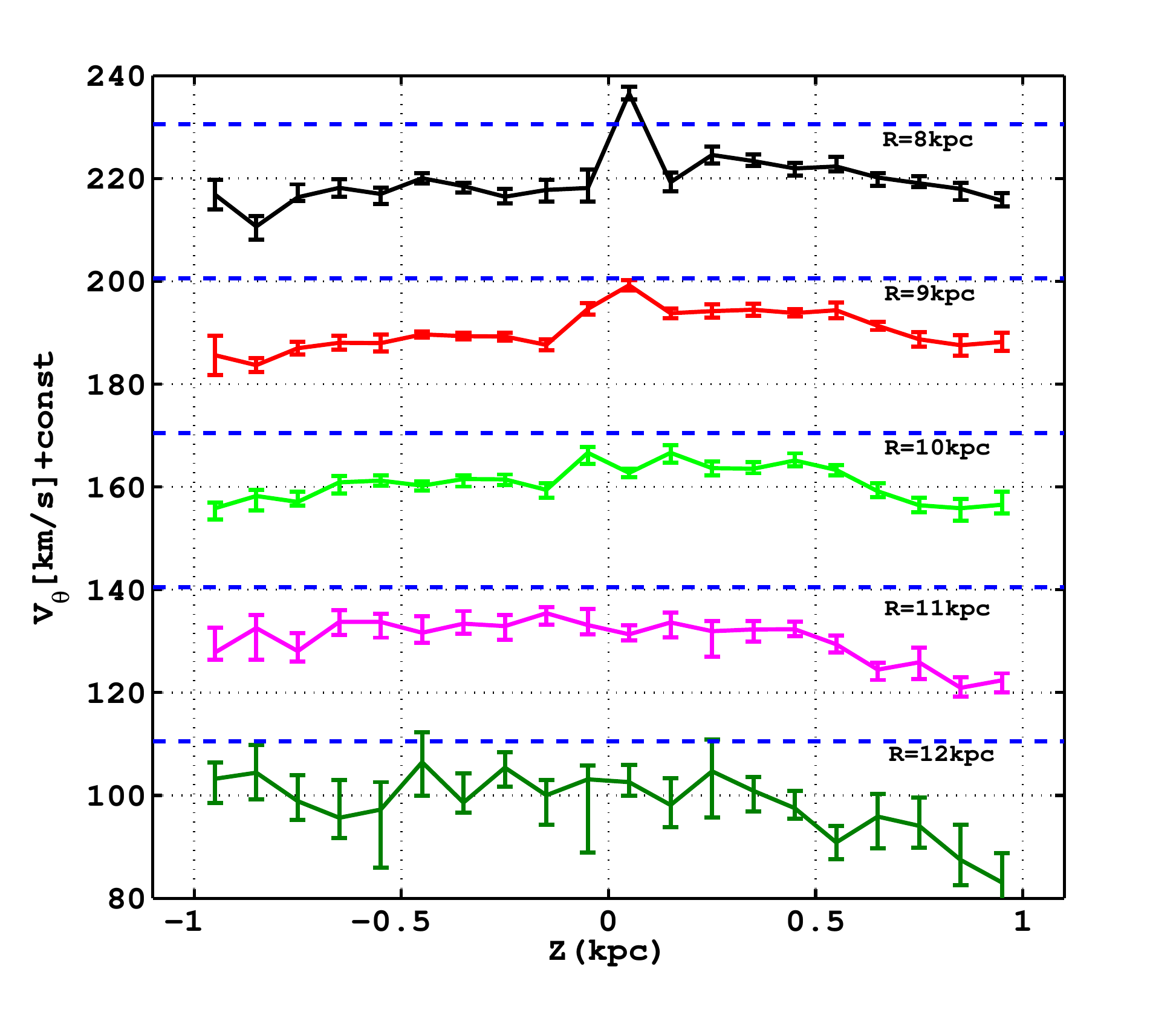}
  \caption{Similar to Fig.~\ref{1DVPHI}, but for azimuthal velocity of stars with [Fe/H]$>$ -0.4 dex and $Z$=[-1, 1] kpc. The variation of azimuthal velocity at radii 8-12 kpc still exists: the stars in the south are rotating faster than those in the north when $|Z|$ $>$ 0.7 \,kpc for the 11 and 12 \,kpc slices. Notice that it looks like it moves gradually from asymmetry with $V_{\theta}$ larger above the plane at $R \approx$ 8 $-$ 9 kpc, to $V_{\theta}$ larger below the plane at $R \approx$ 11 $-$ 12 \,kpc.}
  \label{1DVPHI_test2}
\end{figure}

In \citet{Tian172}, an interesting U-shaped profile is found showing different trends in $V_{\theta}$ with Galactocentric radius. In the region of $R \lesssim$ 10.5 kpc, the mean azimuthal velocities for both the young and old red clump stars  mildly decrease by $\sim$10 km s$^{-1}$. Beyond R $\sim$ 10.5 \,kpc, the mean azimuthal velocities for both populations increase to 240 $-$ 245 km s$^{-1}$ at R $\sim$13 \,kpc. We do not detect similar structure in our K giant sample (see middle panel of Fig.~\ref{1DRVRVPHIVZ}) possibly due to sampling different stellar populations, or using a different deprojection method. Our general trend among lower region stars is similar to \citet{Lop142} for $R$ less than 13 \,kpc. As mentioned, we find that there is an asymmetry at $R  \approx $ 8 $-$ 9  \, kpc, at least for $Z \lesssim$ 0.5 \,kpc and $R \sim$ 10-11 \,kpc. We do a simple test for the azimuthal velocity distribution by limiting the sample to  [Fe/H]$>$ -0.4 dex and $Z$ = [-1, 1] \,kpc to see more details of the midplane and test whether there are still clear disk asymmetries and variations. We can see in Fig.~\ref{1DVPHI_test2}, at $R=$ 11-12 \,kpc, that stars on the south side are still faster than north side when $|Z|$ $ >$ 0.7 \, kpc, and notice that it looks like it moves gradually from asymmetry with $V_{\theta}$ larger above the plane at $R \approx$ 8 $-$ 9 kpc, to $V_{\theta}$ larger below the plane at $R$ = 11 $-$ 12 \,kpc.  Recently, a southern rising trend of $V_{\theta}$ in the range $Z$ = [-2, 2] \,kpc, $R$ = [8, 10] \,kpc is also found by \citet{Pearl17} using LAMOST DR3 F-type stars. This may be a real effect that merits future validation with higher precision proper motions and a larger sample. Some similar possible mechanisms have been introduced in the radial velocity section, whose theoretical details on the mechanisms are a task for a future work.

The bottom panel of Fig.~\ref{1DRVRVPHIVZ} shows that the median $V_{Z}$ varies from negative to positive with radius from inside to outside $R_\odot$. Recently, Fig. 14 of  \citet{Carrillo17} displayed a clear negative gradient above the plane inside the solar location. The positive gradient outside the solar radius is similar to that found by \citet{Liu172} in the range of $R$ less than 12 \,kpc, which showed a significant bulk motion for lower old red clump stars. The authors suggested that it is hard to explain their results by the perturbation due to a merging event \citep{Gomez13} or spiral arms \citep{Debattista99}, because their measured vertical velocity is too large compared to perturbations seen in the models. \citet{Liu172} suggested that the kinematics result from warp line-of-node kinematics. But our measured velocity (see the bottom panel of Fig.~\ref{1DRVRVPHIVZ}) is similar to that predicted by the minor merger simulation of \citet{Gomez13}. Their results displayed variations of vertical motions less than $\sim10$ km s$ ^{-1} $. They also predicted radial and azimuthal variations of the mean vertical velocity, correlating with the spatial structure as mentioned previously. We think that spiral arms are not likely important for vertical motions in the outer disk. \citet{Monari15} pointed out the Galactic bar is unlikely to induce mean vertical motions greater than $\sim$ 0.5 km s$^{-1}$ in the outer disk; our bulk motions are larger than 0.5 km s$^{-1} $, so we do not favor a bar/spiral arm explanation for the vertical asymmetric motions. A minor merger or other external perturbation might be the most reasonable explanation. However, we can not exclude a warp contribution, as suggested by \citet{Lop141}, or that from the dark matter halo. 

In summary, by combining our measured asymmetrical distributions and comparisons with other works, we conclude that planar asymmetrical motions are not likely mainly contributed by the spiral arms or minor mergers, and vertical asymmetrical motions are not likely mainly contributed by the bar or spiral arms. Other mechanisms such as the warps, the dark matter halo, etc. are needed to be considered. In the future, we will use Gaia DR2 and test particle simulations to validate these results.

\subsection{Reconstructing previous LAMOST work}
\begin{figure}
  \centering
  \includegraphics[width=0.52\textwidth]{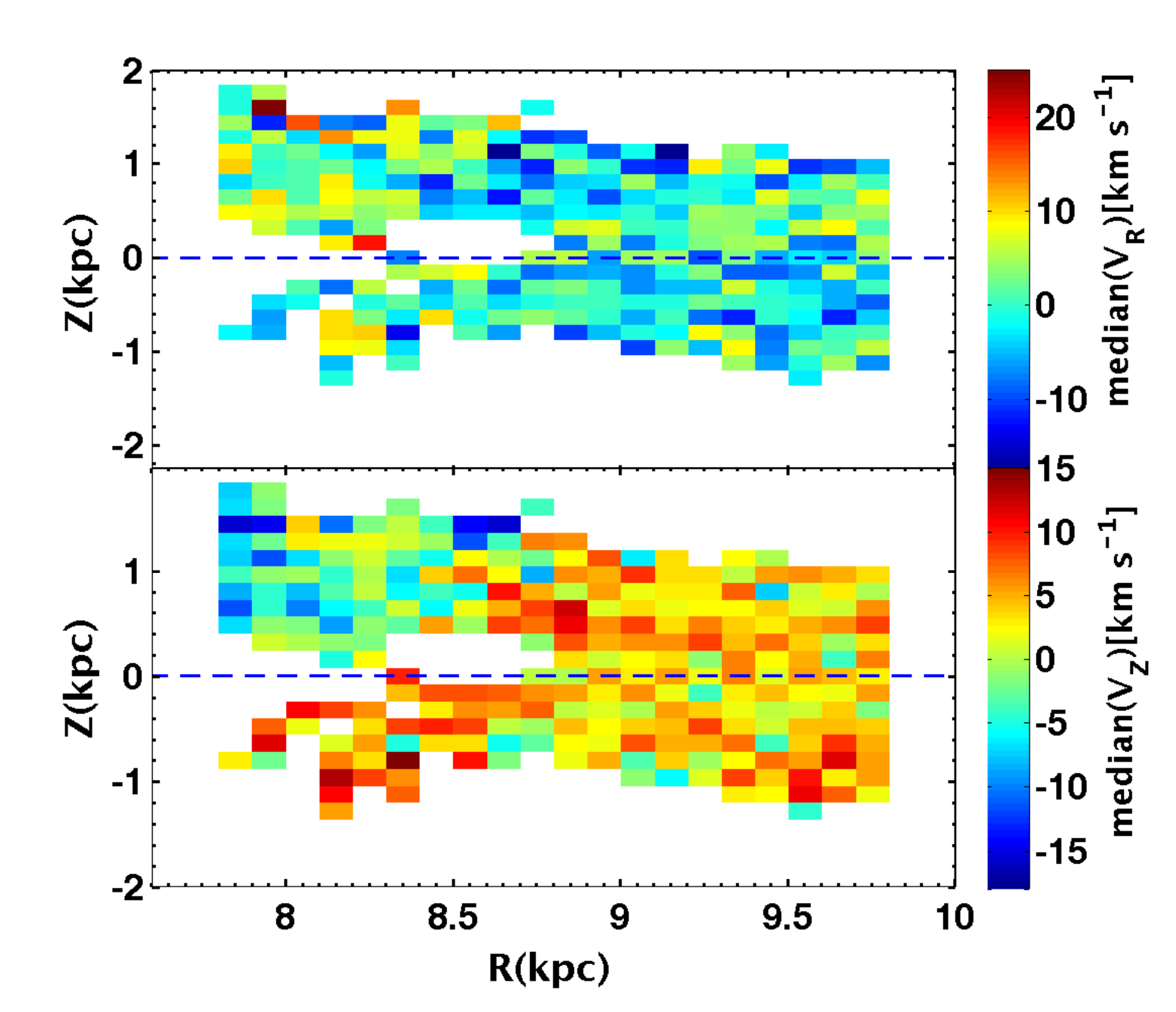}
  \caption{Similar to the top panel of Fig.~\ref{RZVR}, but for vertical and radial velocity on the $R, Z$ plane. The sample is limited to $Z$=[-2, 2] \,kpc and $R$=[7.8, 9.8] \,kpc, in order to reconstruct and compare with earlier LAMOST work by  \citet{Carlin13}.}.
  \label{RZVRVZ_Carlin}
\end{figure}

Figure~\ref{RZVRVZ_Carlin} shows the binned median radial and vertical velocities $V_{R}$ and $V_{Z}$ in the $R, Z$ plane in the range $R$ = 7.8-9.8 \,kpc, $Z$ = -2 to 2 \,kpc. Although we do not have as many K giant stars with proper motion as the 400000 F-stars used by \citet{Carlin13} (hereafter, Carlin13), and the spatial coverage is worse, we directly compare with Carlin13 for corresponding regions that are well sampled. The upper panel of Fig.~\ref{RZVRVZ_Carlin} shows that the $V_{R}$ trend has some differences in the north with respect to the Carlin13 work. We measure median $V_{R}$ in the range $R\approx $ 8.6 to 9.2 \,kpc to be negative (on average), while Carlin13 found $V_R$ in this region to be around zero, and furthermore did not see the oscillation that we find in Fig.~\ref{RZVR}. The difference might be caused by the lower precision of the PPMXL proper motions used in that work, as well as differences in the distance determination methods. \citet{Carrillo17} showed that proper motion uncertainty and distance error can lead to different or reversed asymmetrical motions in the solar neighborhood by reconsidering the bending or breathing mode with Radial Velocity Experiment (RAVE) latest sample \citep{Kunder17} and the Tycho-Gaia astrometric solution catalogue (TGAS) \citep{Gaia161}. Other factors such as the solar motion, solar location, selection criteria of different tracers, and the adopted local standard of rest velocity can also produce some differences. In the south, for $V_{R}$, the overall trends are similar in our work to those of Carlin13; many bins are less than or around 0 km s$^{-1}$ with some positive bins mixed in. There is little difference between the $V_{Z}$ trends inside the solar radius; above the plane the velocity is moving downward on average, while below the plane, the velocities are moving upward. However, outside the Sun, we find upward bulk vertical motions, while Carlin13 found that the compression-like motions continued to beyond the solar radius.

The Carlin13 work has also been revisited by \citet{Pearl17} with new corrections to the PPMXL zero points; their results are very similar to what we have found. They also do not find substantially positive $V_{R}$ at high $Z$ due to elimination of data with larger errors, and they do not observe the north region stars moving down toward the plane similar to our work. \citet{Pearl17} suggest that the difference arises because the Carlin13 sample contained an overwhelming amount of data in the northern third quadrant. These differences are clearer for us to see between Fig.~\ref{RZVRVZ_Carlin} in this work and Fig. 2 in \citet{Carlin13}.

\section{Conclusions}
In this work, we investigate the kinematics of K giant stars in the Galactic disk between $R$ = 6 to 13 \,kpc, $Z$ = -2 to 2 \,kpc with the GPS1 proper motion catalogue. From asymmetrical variations of the velocities with Galactocentric radius and height for 65000 K giant or RGB stars, we conclude the following:

The median radial velocity profile $V_R$ has a large northern oscillating structure from $R=$6 \,kpc to 13 \,kpc. There is velocity substructure located at $Z\sim$ 0.5 \,kpc and $R\sim$ 10$-$11 \,kpc for the $V_R$ distribution corresponding to the north $-$ near overdensity we have confirmed previously.

We discover asymmetrical rotational motions when comparing stars above and below the plane at $R \geq$ 10 \,kpc and $|Z| \geq$ 0.5 kpc. Stars in the south are, on average, rotating faster than those in the north. At $R  \approx $ 8 $-$ 9  \,kpc, $|Z| \lesssim$ 0.5 \,kpc, we show that the northern stars are rotating faster than their counterparts in the south.

We also find that there is a compression motion at radii inside the solar location, and upward asymmetrical bulk motions outside the solar radius until $\sim$ 13 \,kpc.
The radial profile of median $V_{Z}$ displays a trend from downward to upward bulk motions in the range of 6 to 13 \,kpc, with all regions in our study beyond $R \gtrsim 9$~kpc moving upward on average.
  
With the help of previous works, we discuss that in$-$plane asymmetries are not mainly contributed by the spiral arms or minor merger, and because we see vertical asymmetries, we can rule out that the vertical features are mainly caused by spiral arms or the bar. We cannot exclude contributions such as the warps, the dark matter halo, or other mechanisms. It will require more detailed mapping of disk velocity substructure(s) to decipher the mechanisms creating these structures in the future.

This paper can be thought of as a pilot paper for outer disk kinematics. ESA's mission Gaia \citep{Gaia161} will soon release highly accurate parallaxes and proper motions for over a billion sources, which will provide an opportunity to map the outer disk in more detail and hopefully answer some of puzzles presented here. However, fully exploiting the proper motions from Gaia for disk kinematics requires a large spectroscopic sample such as LAMOST to provide the radial velocities and stellar parameters.

\section*{Acknowledgements}
We would like to thank the anonymous referee for his/her helpful comments. We also thank Liu C., Lawrence M. Widrow, Francesca Figueras, Alice C. Quillen, Ismael E. Carrillo, and Ivan Minchev for helpful discussions and comments. This work was supported by the National Key Research and Development Program of China through grant 2017YFA0402702.
MLC was supported by grant AYA2015-66506-P of the Spanish Ministry of Economy and
Competitiveness. Guoshoujing Telescope (the Large Sky Area Multi-Object Fiber Spectroscopic Telescope LAMOST) is a National Major Scientific Project built by the Chinese Academy of Sciences. Funding for the project has been provided by the National Development and Reform Commission. LAMOST is operated and managed by the National Astronomical Observatories, Chinese Academy of Sciences.

\label{lastpage}
\end{document}